EXAMINING THE LINK BETWEEN PEROXIREDOXIN PROTEINS AND MUTUALLY EXCLUSIVE TRANSCRIPTION FACTOR ACTIVATION WITH A MATHEMATICAL MODEL

By

ZACHARY ALLEN SCHLAMOWITZ

A Thesis Submitted to The W.A. Franke Honors College

In Partial Fulfillment of the bachelor's degree
With Honors in

Mathematics

THE UNIVERSITY OF ARIZONA

M A Y 2 0 2 3

Approved by:

Dr. Kevin Lin[1]
Dr. Andrew Paek[2]
[1] Department of Mathematics, [2] Department of Molecular and Cellular Biology

# Examining the link between peroxiredoxin proteins and mutually exclusive transcription factor activation with a mathematical model

Zachary Schlamowitz[1]

[1] Department of Mathematics, University of Arizona

**Abstract**

Oxidative stress is a fundamental stimulus to which eukaryotic cells respond via many channels. Among these channels are both protein systems that process oxidative stress, such as the 2-Cys peroxiredoxin-thioredoxin system (PTRS), as well as changes in transcriptional activity that target outcomes such as growth, damage control and repair, or cell death. Recent work has revealed connections between the PTRS and temporal phases of transcriptional activity involving famous transcription factors like p53 and FOXO1. To examine potential mechanisms for these connections, we implement an existing dynamical systems mathematical model for the PTRS. We hypothesize that dose-dependent hyperoxidation events enact ultrasensitive switches by which the PTRS can categorize stress severity and activate appropriate transcriptional responses. Using numerical simulations of the PTRS in human cells, we provide a proof of principle for staggered, switchlike hyperoxidation of peroxiredoxins (Prx) as well as an underlying mechanism requiring simultaneous signaling by Prx I and II. Then, we use our model to make testable predictions about individual Prx knockouts as well as the affinity for hydrogen peroxide of Prx across oxidation states. This study provides direction for future experimental work and sheds light into the mechanisms underlying oxidative stress response in human cells.

## 1 Introduction

Oxidative stress is a prominent stress faced by all cell types caused by the presence of reactive oxygen species (ROS) inside or around the cell. Hydrogen peroxide ($H_2O_2$) is a prominent ROS which engages in complex interactions with cells. Depending on the level of $H_2O_2$ exposure, cells can engage widely varying routines, ranging from cell growth to stress response mechanisms. Low levels of $H_2O_2$ induce eustress, mild oxidative stress that is sometimes considered beneficial and can encourage cells to proliferate or differentiate for such outcomes as wound healing[1]-[6]. $H_2O_2$ often elicits such responses by acting as a signaling molecule, activating various signal transduction pathways in the cell[7]-[8]. However, high levels of $H_2O_2$ induce distress, defined as oxidative stress which is toxic to the cell. Distress can cause DNA damage and protein misfolding and can lead to growth arrest and cell death on the cellular level [9]-[10]. Such divergent responses to eustress and distress are believed to involve the activation of differing sets of transcription factors; recent work from Jose et al. (2023) supports this hypothesis and has characterized certain transcription factors associated with particular stress response regimes[11].

Among these differentially activated transcription factors are p53 and FOXO1 proteins. Famous for their roles in regulating cell fate outcomes such as cell cycle arrest and apoptosis, these proteins can also activate NADPH/GSH production, ROS scavenger enzymes, DNA damage repair, autophagy, and protein quality control routines[12]-[20]. As such, it is unsurprising that both transcription factors would be involved in oxidative stress responses. However, recent evidence suggests that the dynamics of their involvement are counterintuitive; Jose et al. show that following exposure to $H_2O_2$ at doses which are known to induce DNA damage (a major activator of p53), human breast cancer cells exhibit a temporally biphasic response in which initially FOXO1 is activated while p53 is suppressed; subsequently, p53 is activated, but only once FOXO1 transcriptional activity is suppressed. Furthermore, high doses of $H_2O_2$ result only in the FOXO1 phase of transcriptional activity, with cell death ensuing while p53 remains suppressed. Moreover, the transition between these regimes is relatively sharp in time, assuming a switchlike shape. Finally, the duration of the first phase of this response is $H_2O_2$-dose dependent, suggesting the $H_2O_2$ dose information is in some capacity internalized[11]. Consequently, it is natural to wonder by what mechanism(s) do cells categorize oxidative stress into varying responses, and why might they do so.

A key regulator of intracellular $H_2O_2$ concentrations is the cytoplasmic 2-Cys peroxiredoxin-thioredoxin system (PTRS). Consisting of 2-Cys peroxiredoxins (Prx), thioredoxin (Trx), thioredoxin reductase (TrxR), and sulfiredoxin (Srx), the system is integral to cytoplasmic redox biology in eukaryotes, including redox signaling as well as protection against and management of oxidative stress. To mitigate oxidative stress, the PTRS cycles oxidation states across peroxiredoxins, using thioredoxins to safely dispose of the extra oxidation. Core to the functionality of the PTRS are the cysteine residues of the peroxiredoxins; each Prx has two such residues which must both become oxidized to allow transfer of the gained electrons to thioredoxins. Under normal functioning of the system (in response to low $H_2O_2$ concentrations), the gained electrons are captured in the form of a disulfide bond that forms as peroxiredoxins dimerize after being oxidized, and it is this disulfide bond that is transferred to thioredoxin. It is believed that downstream transfer of the disulfide bond may be a key mechanism for redox signaling. However, high concentrations of $H_2O_2$ can cause hyperoxidation of the peroxiredoxins (double or triple oxidation of Prx-SH to Prx-$SO_{2/3}$). Although this hyperoxidation is mitigated via reduction by sulfiredoxin, hyperoxidation of peroxiredoxins disables their functionality and leads to aggregation of the hyperoxidized proteins. Significant hyperoxidation occurs once $H_2O_2$ concentrations exceed a particular threshold[21].

Jose et al. implicate the PTRS in their study of the biphasic regimes of transcriptional activity following oxidative stress, as they are able to experimentally change the duration of the phases via modulation of peroxiredoxin hyperoxidation states[11]. Such a connection is intuitive, given the PTRS's prominent roles in redox signaling and oxidative stress response. However, links between the PTRS and specific transcription factor dynamics are not well understood. Nevertheless, work in redox biochemistry has revealed suggestive properties of the PTRS itself [22]. There are six different species of peroxiredoxin within the 2-Cys peroxiredoxin family; of these, PrxI, PrxII, and PrxV are of note, as Bolduc et al. (2018) show that they have significantly different $H_2O_2$ concentration thresholds for hyperoxidation, demarking sharp transitions; Portillo-Ledesma et al. (2018) show a similar result for oxidation thresholds as well[21]-[23]. We believe that hyperoxidation is ultrasensitive in dose, so that in the presence of $H_2O_2$ concentrations above its hyperoxidation threshold, a given peroxiredoxin becomes predominantly hyperoxidized, whereas for $H_2O_2$ concentrations below the threshold it does not. Since hyperoxidation disables the peroxiredoxins, these hyperoxidation dose thresholds can effectively establish activation/deactivation switches for PrxI, II, and V. Such switches may be relevant to understanding the switch-like temporal phases in cellular responses to oxidative stress.

To examine this possibility, we turn to a computational proof-of-principle approach. Recent work has seen related mathematical modeling of $H_2O_2$ signaling, individual peroxiredoxin activity, and the PTRS as a whole[24]-[26]. In a comprehensive analysis, Selvaggio et al. (2018) mathematically model the PTRS with nonlinear ordinary differential equations (ODEs) describing changes in the concentrations of the protein components of the system[24]. Using parameters drawn from 14 different cell lines, they apply design space analysis to identify all possible qualitative outcomes of the modeled system. While their initial model only describes one peroxiredoxin species, Selvaggio et al. propose a second version of their model accounting for both two distinct peroxiredoxin species (Prx1 and Prx2) as well as $H_2O_2$ membrane permeability[24]. We take this second model as the foundation of our theoretical examination of temporally biphasic peroxiredoxin hyperoxidation and its possible implications for cellular regimes.

We hypothesize that peroxiredoxins PrxII and PrxI function as oxidative stress sensors. Specifically, we suspect that the switchlike hyperoxidation thresholds of PrxII and PrxI recognize and categorize oxidative stress into varying degrees of severity, and correspondingly activate appropriate transcriptional regimes (e.g., pro-growth regimes for eustress, damage control regimes for middling stress, and high-priority damage repair for distress). We suppose these categorizations are communicated for transcription by the transfer of the peroxiredoxin's disulfide bonds, perhaps through complex redox signal transduction pathways. In this work, assume that the disulfide state is the signaling state of the PTRS. Our examination is focused on examining two possible mechanisms for the encoding of stress levels by the PTRS and their activation of transcriptional regimes; these are summarized in Figure 1. Hypothesis I supposes that disulfide signaling activity of each peroxiredoxin corresponds individually to each of the phases of transcriptional activity observed by Jose et al.; e.g., PrxII activity alone corresponds to a growth phase, while PrxI activity alone corresponds to the p53 phase. Contrastingly, hypothesis II supposes that it is combinations of peroxiredoxin signaling activity that activate transcriptional regimes; e.g., activity of PrxII, PrxI, and PrxV corresponds to the growth phase while activity of just PrxI and PrxV corresponds to the p53 phase. These hypothesized mechanisms are not exhaustive and may not be fully mutually exclusive but provide a guiding framework within which we approach our model simulations.

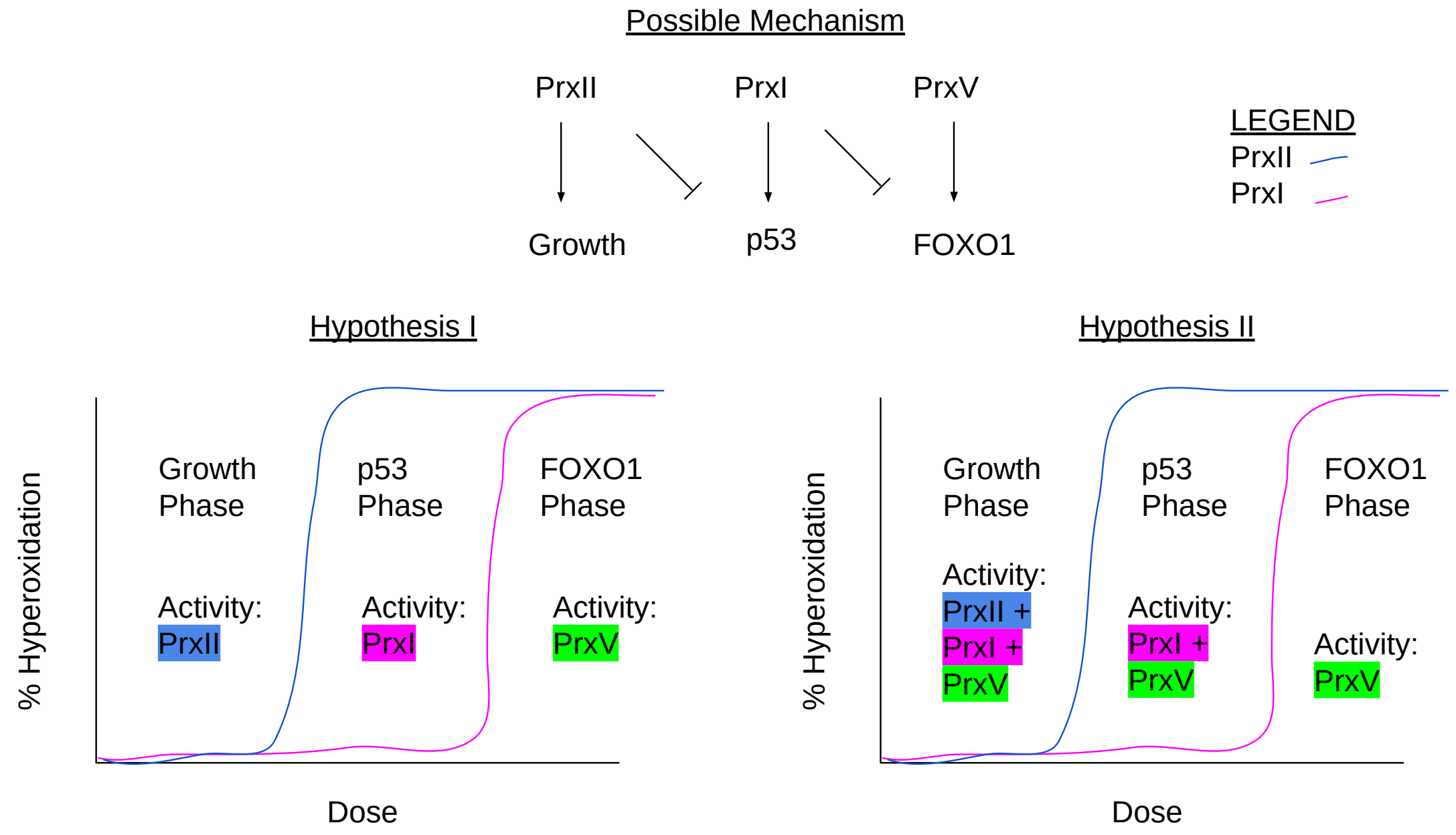

**Figure 1:** Competing hypotheses describing staggered ultrasensitive peroxiredoxin hyperoxidation switches and mechanisms of activation for them. Hypothesis I proposes isolated peroxiredoxin signaling activity as responsible for temporal phases of transcriptional activity, whereas hypothesis II proposes layered co-activity of peroxiredoxin signaling.

## 2 Methods

### *2.1 Model Structure*

The two-species model of the PTRS put forth by Selvaggio et al. is a system of nine nonlinear ordinary differential equations (ODEs) (Figure 2). These equations describe the temporal dynamics of the PTRS and can be grouped into four categories: those pertaining to (1) hydrogen peroxide, (2) peroxiredoxin I, (3) peroxiredoxin II, and (4) thioredoxin. Equations (1)-(2) are devoted to hydrogen peroxide dynamics; equation (1) describes an extracellular $H_2O_2$ compartment, which allows movement through membrane permeability into an intracellular compartment, the dynamics of which are given in equation (2). Peroxiredoxins are simplified into four chemical states: thiol Prx-$S^-$, sulfenic/oxidized Prx-$SO^-$, sulfinic/sulfonic/hyperoxidized Prx-$SO_{2/3}^-$, and disulfide Prx-SS. As these states are related by a total cytosolic Prx concentration, only three of them need to be modeled; without loss of generality, Prx-$SO^-$, Prx-$SO_{2/3}^-$, and Prx-SS are chosen. Equations (3)-(5) give the dynamics of these compartments for PrxI while (6)-(8) do the same for PrxII. In similar fashion, equation (9) describes the dynamics of Trx-SS, from which all Trx states can be deduced. Finally, equations (10)-(12), which are not ODEs, give the rules by which the Prx and Trx compartments are related.

(a)

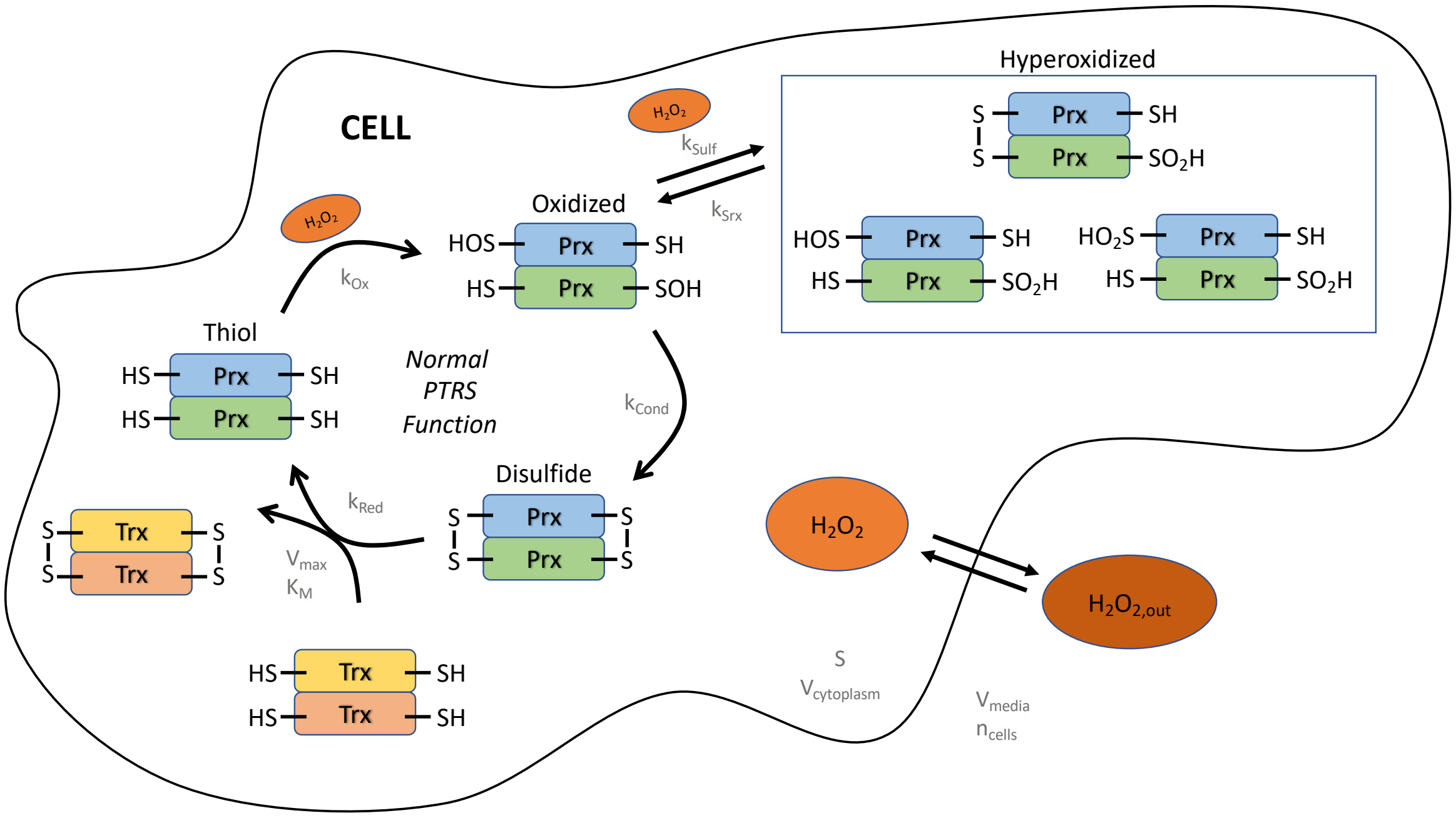

(b)

$$\frac{dH_2O_{2,out}}{dt} = \frac{\kappa n_{cells} S}{V_{media}}(H_2O_2 - H_2O_{2,out}) \tag{1}$$

$$\begin{aligned}\frac{dH_2O_2}{dt} &= \frac{\kappa S}{V_{cytoplasm}}(H_2O_{2,out} - H_2O_2) \\ &\quad - k_{Alt}H_2O_2 - k_{Ox}(PrxI\text{-}S^- + PrxII\text{-}S^-)H_2O_2 \\ &\quad - k^I_{Sulf}PrxI\text{-}SO^- H_2O_2 - k^{II}_{Sulf}PrxII\text{-}SO^- H_2O_2\end{aligned} \tag{2}$$

$$\begin{aligned}\frac{dPrxI\text{-}SO^-}{dt} &= k_{Ox}PrxI\text{-}S^- H_2O_2 + k_{Srx}PrxI\text{-}SO_2^- \\ &\quad - k^I_{Sulf}PrxI\text{-}SO^- H_2O_2 - k^I_{Cond}PrxI\text{-}SO^-\end{aligned} \tag{3}$$

$$\frac{dPrxI\text{-}SO_2^-}{dt} = k^I_{Sulf}PrxI\text{-}SO^- H_2O_2 - k_{Srx}PrxI\text{-}SO_2^- \tag{4}$$

$$\frac{dPrxI\text{-}SS}{dt} = k^I_{Cond}PrxI\text{-}SO^- - k_{Red}Trx\text{-}S^- PrxI\text{-}SS \tag{5}$$

$$\begin{aligned}\frac{dPrxII\text{-}SO^-}{dt} &= k_{Ox}PrxII\text{-}S^- H_2O_2 + k_{Srx}PrxII\text{-}SO_2^- \\ &\quad - k^{II}_{Sulf}PrxII\text{-}SO^- H_2O_2 - k^{II}_{Cond}PrxII\text{-}SO^-\end{aligned} \tag{6}$$

$$\frac{dPrxII\text{-}SO_2^-}{dt} = k^{II}_{Sulf}PrxII\text{-}SO^- H_2O_2 - k_{Srx}PrxII\text{-}SO_2^- \tag{7}$$

$$\frac{dPrxII\text{-}SS}{dt} = k^{II}_{Cond}PrxII\text{-}SO^- - k_{Red}Trx\text{-}S^- PrxII\text{-}SS \tag{8}$$

$$\frac{dTrx\text{-}SS}{dt} = k_{Red}Trx\text{-}S^-(PrxI\text{-}SS + PrxII\text{-}SS) - \frac{V^{App}_{Max}Trx\text{-}SS}{K_M + Trx\text{-}SS} \tag{9}$$

$$PrxI_{Total} = PrxI\text{-}S^- + PrxI\text{-}SS + PrxI\text{-}SO^- + PrxI\text{-}SO_2^- \tag{10}$$

$$PrxII_{Total} = PrxII\text{-}S^- + PrxII\text{-}SS + PrxII\text{-}SO^- + PrxII\text{-}SO_2^- \tag{11}$$

$$Trx_{Total} = Trx\text{-}S^- + Trx\text{-}SS \tag{12}$$

**Figure 2:** The PTRS system and the mathematical model. (a) Schematic of PTRS. (b) the two-peroxiredoxin ODE model proposed by Selvaggio et al. (2018)

*2.2 Technical Details*
We designed a computational pipeline in MATLAB (using versions R2022a, R2022b, and R2023a across devices) to simulate the PTRS. To implement the two-peroxiredoxin model which separately specifies PrxI and PrxII (Selvaggio et al., Supplementary Information, p.52), we explored a variety of proprietary MATLAB differential equation solvers[24]. (Selvaggio et al.'s work was performed in Mathematica, meaning our MATLAB implementation is entirely separate.) Solvers for non-stiff ODEs, such as **ode23**, struggled to find solutions, revealing that our model contains a stiff system of ODEs. As such, we used the stiff solver variants such as **ode23s** in the implementation, which worked without issue. The code in its entirety is available on Github and can be accessed publicly at https://github.com/ZachSchlamowitz/OxStress. However, the core structure of the code and its key functionalities are summarized below in Figure 3.

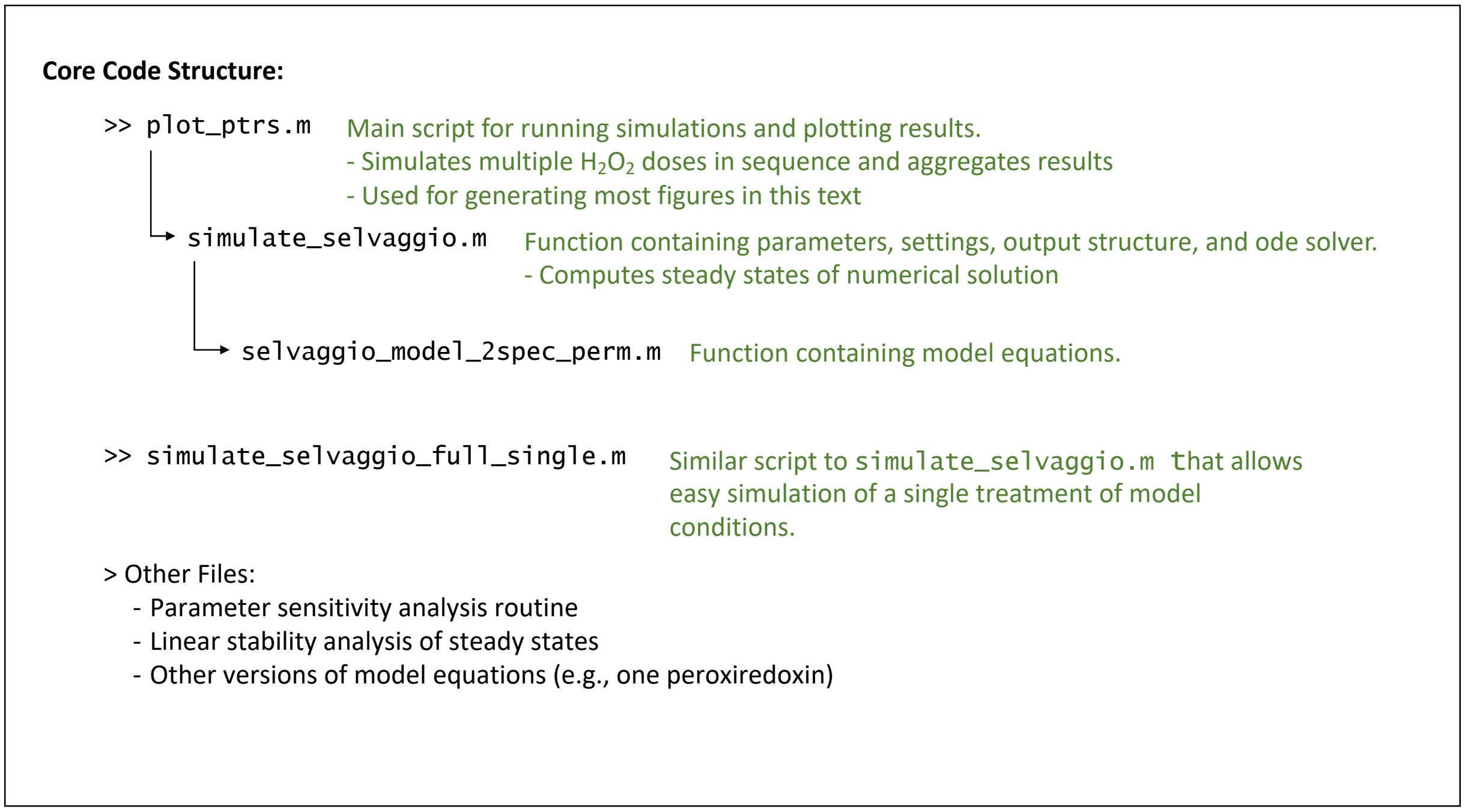

**Figure 3:** Key files of the coding pipeline and descriptions of the functionality.

*2.3 Parameters*

Parameters for the two-peroxiredoxin model are presented in Table 1 for two cell types: MCF-7 human breast cancer cells and HEK293 human kidney cells. Selvaggio et al. examine parameters for 14 cell lines, but here we simplify this to focus only on these two, selected for the following reasons. MCF-7 cells are available for proprietary experiments and were used in the motivating work by Jose et al.[11]. However, Supplementary Figures S9 and S11 of Selvaggio et al. use HEK293 parameters[24]. In order to replicate these results, we implemented HEK293 parameters as well. All simulation results in this work are qualitatively the same for both cell lines; however, the results using HEK293 parameters show improved clarity and, as HEK293 parameters have been more thoroughly verified across experimental work, we have selected HEK293 as our model cell line for the figures shown in this text.

**Table 1:** Model Parameters

| Parameter | MCF-7 | HEK293 |
|---|---|---|
| $k_{Alt}$ ($sec^{-1}$) *rate constant for alternate $H_2O_2$ sinks* [1] | 79 | 160 |
| $k_{Ox}$ ($\mu M^{-1}$ $sec^{-1}$) *rate constant for PrxI/II first oxidation* [1] | 40 | 40 |
| $k_{Srx}$ ($sec^{-1}$) *rate constant for Srx activity (reduction of PrxI/II-$SO_{2/3}$)* [1] | 3.3e-3 | 4.1e-4 |
| $k_{Red}$ ($\mu M^{-1}$ $sec^{-1}$) *rate constant for reduction of PrxI/II-SS to PrxI/II-$S^-$* [1] | 0.21 | 0.21 |
| $V^{App}_{Max}$ (μM/sec) *max rate of TrxR activity* [1] | 230 | 190 |
| $K_M$ (μM) *for TrxR activity* [1] | 1.8 | 1.8 |
| TrxTotal (μM) *total concentration of Trx* [1,2] | 20 | 46 |
| $k_{Sulf}{}^{I}$ ($\mu M^{-1}$ $sec^{-1}$) *rate constant for PrxI hyperoxidation (MCF7[2], HEK293[1])* | 1.5e-3 | 1.3e-3 |
| $k_{Sulf}{}^{II}$ ($\mu M^{-1}$ $sec^{-1}$) *rate constant for PrxII hyperoxidation (MCF7[2], HEK293[1])* | 3.4e-3 | 1.2e-2 |
| $k_{Cond}{}^{I}$ ($sec^{-1}$) *rate constant for PrxI condensation (MCF7[2], HEK293[1])* | 11 | 9 |
| $k_{Cond}{}^{II}$ ($sec^{-1}$) *rate constant for PrxII condensation (MCF7[2], HEK293[1])* | 0.5 | 1.7 |
| PrxITotal (μM) *total concentration of PrxI (MCF7[2], HEK293[1])* | 110 | 110 |
| PrxIITotal (μM) *total concentration of PrxII (MCF7[2], HEK293[1])* | 30 | 32 |
| $n_{cells}$ (scalar) *number of cells in media / well of plate* [3] | 6000 | 3e5 |
| κ (μm/sec) *permeability coefficient in erythrocytes* [2] | 15 | 15 |
| $V_{media}$ ($\mu m^3$) *volume of media* [6,7] | 2e11 | 2e11 |
| $V_{cytoplasm}$ ($\mu m^3$) *volume of cytoplasm* [4,5] | 1760 | 1150 |
| S ($\mu m^2$) *surface area of cell* [4,5] | 1224.18 | 530 |

Source:
[1] [24], [2] Correspondence with Armindo Salvador (senior author of Selvaggio et al.), [3] [27], [4] [28], [5] [29], [6] [30], [7] Estimate from proprietary experiments

Selvaggio et al. present a comprehensive analysis of the PTRS model, including detailed classification of parameter space into all possible qualitatively different regions and the possible steady states which arise in these regions[24]. As such, performing extensive parameter sensitivity analysis was beyond the scope of this work. However, to check soundness of our numerical solution and to ensure the chosen parameters do not place the PTRS near a bifurcation threshold in parameter space, we performed a simple sensitivity analysis using a scaled $L_2$ norm. This consisted of the following. We first obtained the steady state of the PTRS under the default parameters (Table 1) and under the same parameters save one, which we scaled by a factor of 0.1, 0.5, 2, or 10. We then computed the Euclidean distance between these two steady states (as vectors in $R^9$) and scaled this by the norm of the default steady state vector. Using this algorithm, we proceeded through the parameters one-by-one and confirmed that large (> 1%) changes in the steady state vector only appeared for perturbations of expected parameters: e.g., the total PrxI, PrxII, and Trx concentrations.

*2.4 Sanity Checks*

Prior to running predictive simulations with our model, we applied a variety of tests to confirm proper operation and numerical solution of the model. These included tests that confirmed the correctness of the numerical solution and simulations to replicate results from Selvaggio et al.'s usage of the model. The latter we present in section 2.5 below.

*2.4.1 Correctness of Numerical Solution*

We first considered the potential that solutions obtained by the stiff numerical solver were numerical artifacts as opposed to true solutions. To examine this possibility, we performed a variety of experiments, including changing the error tolerance of the solver, fixing the step size of the solver (**ode23s** uses an adaptive step size, by default), and comparing the solutions produced by different solvers (e.g., **ode23s** vs **ode45s**). In all three cases, changes in the resulting solution were negligible, indicating that a true numerical solution was being obtained by **ode23s**.

*2.5 Replication of figures from Selvaggio et al. (2018)*

Finally, in order to confirm that our implementation of the Selvaggio et al. model agreed with their work, and for the sake of scientific replicability, we ran a selection of simulations similar to those presented in Selvaggio et al. (2018).

The first of these simulations used the one-peroxiredoxin version of the model presented in the main text of Selvaggio et al.[24]. We began with this simpler model for the sake of computational ease. Instead of allowing membrane permeability, the one-peroxiredoxin model accounts for $H_2O_2$ as an exogenous stock that enters the model through a fixed input supply rate, $v_{sup}$. Since $H_2O_2$ is a key determinant of peroxiredoxin activity, changing this parameter can have drastic effects on system behavior. To examine these effects, Selvaggio et al. plot the concentration of each species of the model against orders of magnitude of $v_{sup}$ (see [24] Figure 4, panel I). To replicate part of this figure, we used the numerical solution to the one-peroxiredoxin model computed by **ode23s** to plot steady state concentrations of Prx-$S^-$, Prx-$SO^-$, and Prx-$SO_2^-$ versus values of $v_{sup}$ (Figure 4). Steady state concentrations of each species were obtained by simulating a time-course of the species' concentration and identifying the value after a timepoint beyond which changes in the concentration were insignificant. This simple analysis was performed for 60 values of $v_{sup}$ ranging from $10^{-8}$ to $10^{-2}$. Inspection of our results revealed convincing alignment with the three corresponding curves in Selvaggio et al.'s figure, confirming that our implementation of the one-peroxiredoxin model matched theirs well.

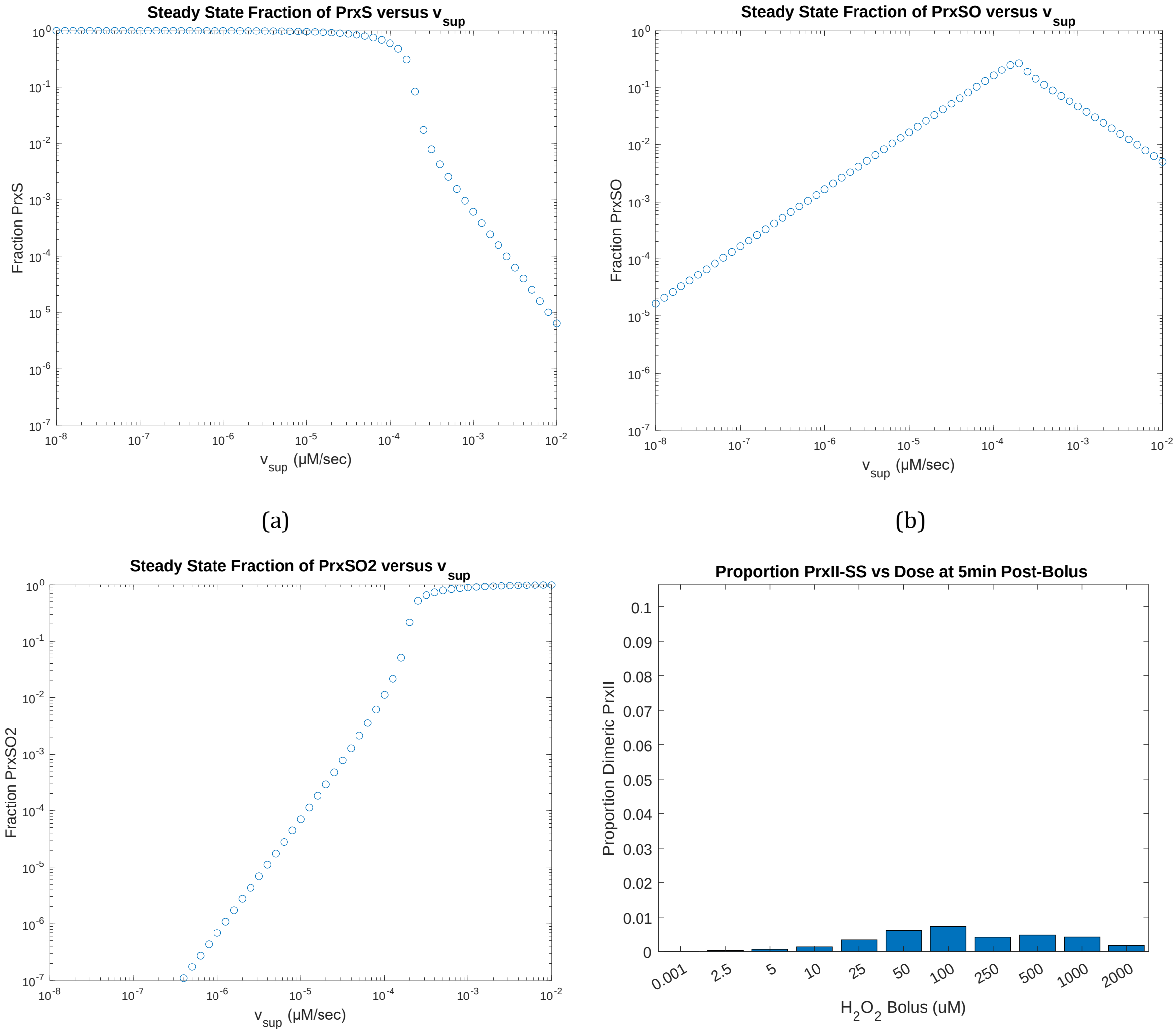

(c) (d)

**Figure 4:** Replication of figures from Selvaggio et al. (2018). Steady state dynamics from the one-peroxiredoxin model are captured in panels (a)-(c) which replicate curves found in panel I of Figure 4 in [24]. Dimeric fractions of PrxII at 5 minutes post-bolus across various bolus doses are shown in panel (d), replicating Supplementary Figure 9B of [24].

Replication results using the two-peroxiredoxin model were less straightforward. We attempted to replicate two figures using the more complex model and only found strong success for one of them. Replication of time series for the fractional activity of PrxI-SS following various bolus doses ([24] Supplementary Figure 11A-E) was very successful, as can be seen by comparison with Figure 7 below. Surprisingly, replication of fractional activity of PrxII-SS at a fixed timepoint following various bolus doses ([24] Supplementary Figure 9B) was less successful. the trend of the dimeric fractions evolves nearly identically in Figure 4d as in Supplementary Figure 9B; oddly however, using the appropriate fractional values (PrxII-SS/PrxIITotal) results in a discrepancy of scale, whereas using gross PrxII-SS values mostly agrees with Supplementary Figure 9B. Even so, this misalignment is somewhat lessened in significance upon further examination of Supplementary Figure S9B. This figure was intended to replicate experimental results presented in Sobotta et al.

(2013) Figure 6B, but comparison of the experimental figure with that of Selvaggio et al. reveals more discrepancies; dimeric PrxII fractions in the Sobotta et al. figure achieve far greater values than in those of the Selvaggio et al. figure, and the trend in dose is somewhat misaligned[24],[27].

Moreover, the larger two-species model inherently gains complexity arising from the number of parameters. Some of these, namely those pertaining to $H_2O_2$ permeability, were unavailable to match with Selvaggio et al.'s implementation, potentially explaining the differences in results. Taking both these parameters and the discrepancies between existing experimental and mathematical results into account, we concluded that our implementation of the two-peroxiredoxin model functions properly.

## 3 Results

Before examining specific results, it is important to recall the motivation for modeling the PTRS. We suspect that activity—specifically, the hyperoxidation—of peroxiredoxins is responsible for activating the mutually exclusive p53 and FOXO1 phases of transcriptional activity in response to $H_2O_2$ exposure. In particular, we hypothesize that hyperoxidation of Prx2 and of Prx1 act as sensors for the severity of oxidative stress, which can then enact three regions of cellular transcriptional activity. The three regions are as follows: with neither peroxiredoxin hyperoxidized, transcriptional activity involves pro-growth pathways; once Prx2 becomes hyperoxidized, growth-pathways are shut off in favor of p53 activity; once Prx1 is hyperoxidized in addition, p53 is inhibited and FOXO1 activity begins. Thus, to examine this hypothesis, we first use the PTRS model as a proof-of-principle to show that sequential dose thresholds for hyperoxidation of Prx2 and Prx1 is possible in such a way as would allow biphasic changes in transcriptional activity in response to oxidative stress.

### *3.1 Staggered hyperoxidation thresholds of PrxI and PrxII emerge in silico*

#### *3.1.1 Hyperoxidation dynamics across doses*

To explore many combinations of $H_2O_2$ doses and timepoints, we simulated the dynamics of the PTRS subsequent to various bolus additions of $H_2O_2$ for a minimum of one hour. We first sought to identify *in silico* the hyperoxidation thresholds (i.e., bolus doses at which each peroxiredoxin becomes hyperoxidized) for PrxI and PrxII. To visualize hyperoxidation as a function of both bolus dose and time, we now consider the fraction of a given peroxiredoxin (I or II) that is hyperoxidized at each timepoint and for each bolus dose in a heatmap (Figure 5).

The model clearly demonstrates the presence of staggered hyperoxidation thresholds, with a majority of PrxII becoming hyperoxidized at doses lower than those for PrxI, similar to established hyperoxidation and first oxidation thresholds[21],[23]. Whereas PrxI only exhibits a notable hyperoxidized fraction after boluses of at least 250μM, PrxII shows comparable hyperoxidation at 50μM.

In addition to becoming hyperoxidized at lower doses, PrxII appears to become hyperoxidized faster when compared to PrxI hyperoxidation at the same dose. For instance, compare PrxI and PrxII hyperoxidation time courses for the 2000μM bolus trial. At this extreme $H_2O_2$ dose, both peroxiredoxins exhibit similar levels of hyperoxidation. However, PrxI displays a visible transition to maximum hyperoxidation in the first few seconds of the simulation, whereas the corresponding transition for PrxII is not visible at the scale of plot (Figure 5). While this may simply result from the larger total population of PrxI (i.e., a greater pool of individual proteins takes longer to hyperoxidize), it also may point to another subtle mechanism by which the PTRS correctly responds to the severity of oxidative stress. Note also the PTRS's response to the oxidative overload causing hyperoxidation in the activity of sulfiredoxin (accounted for by $k_{Srx}$ in the model), which reduces the hyperoxidized peroxiredoxins. This can be seen in the progressive lightening of the time courses with time.

(a)

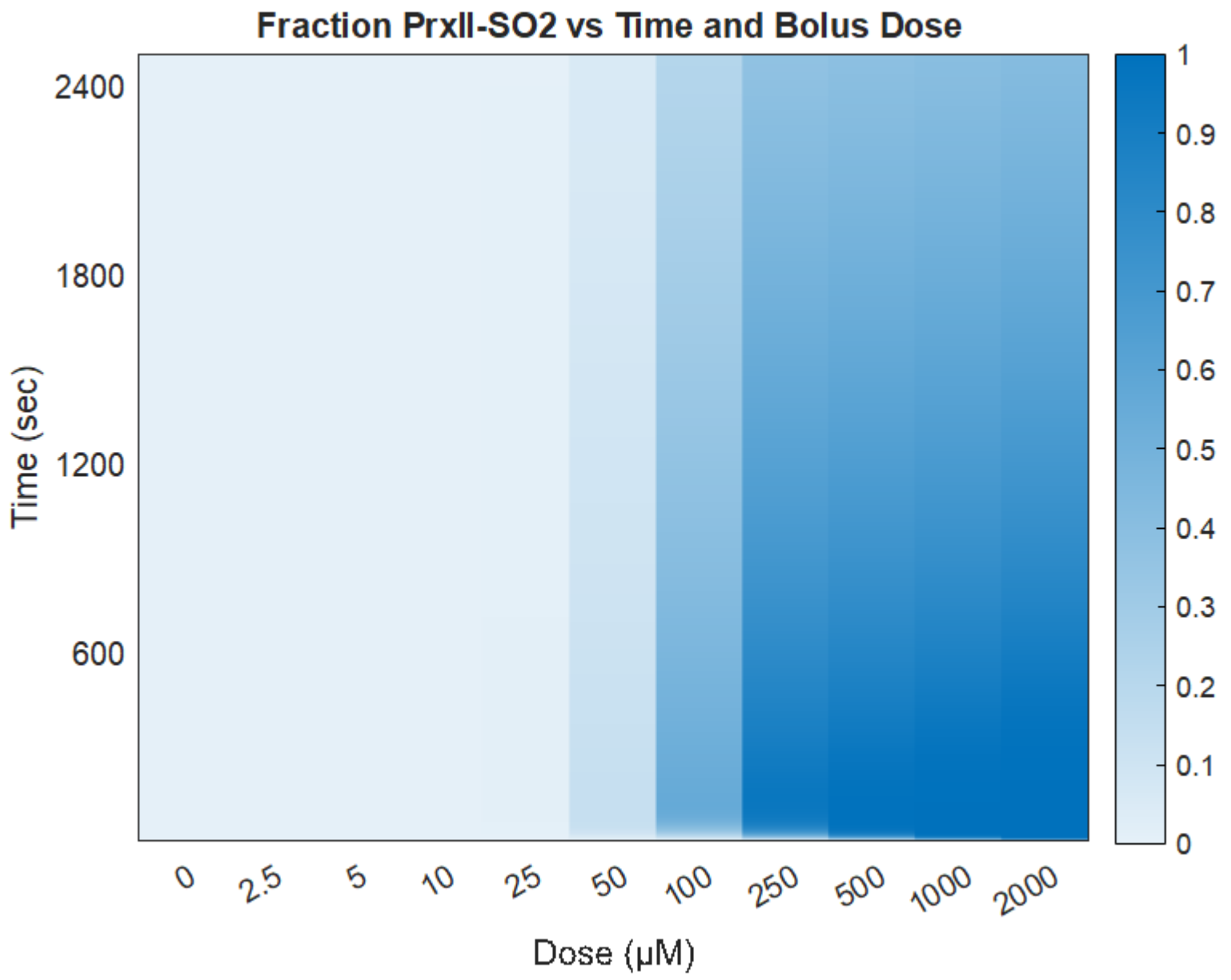

(b)

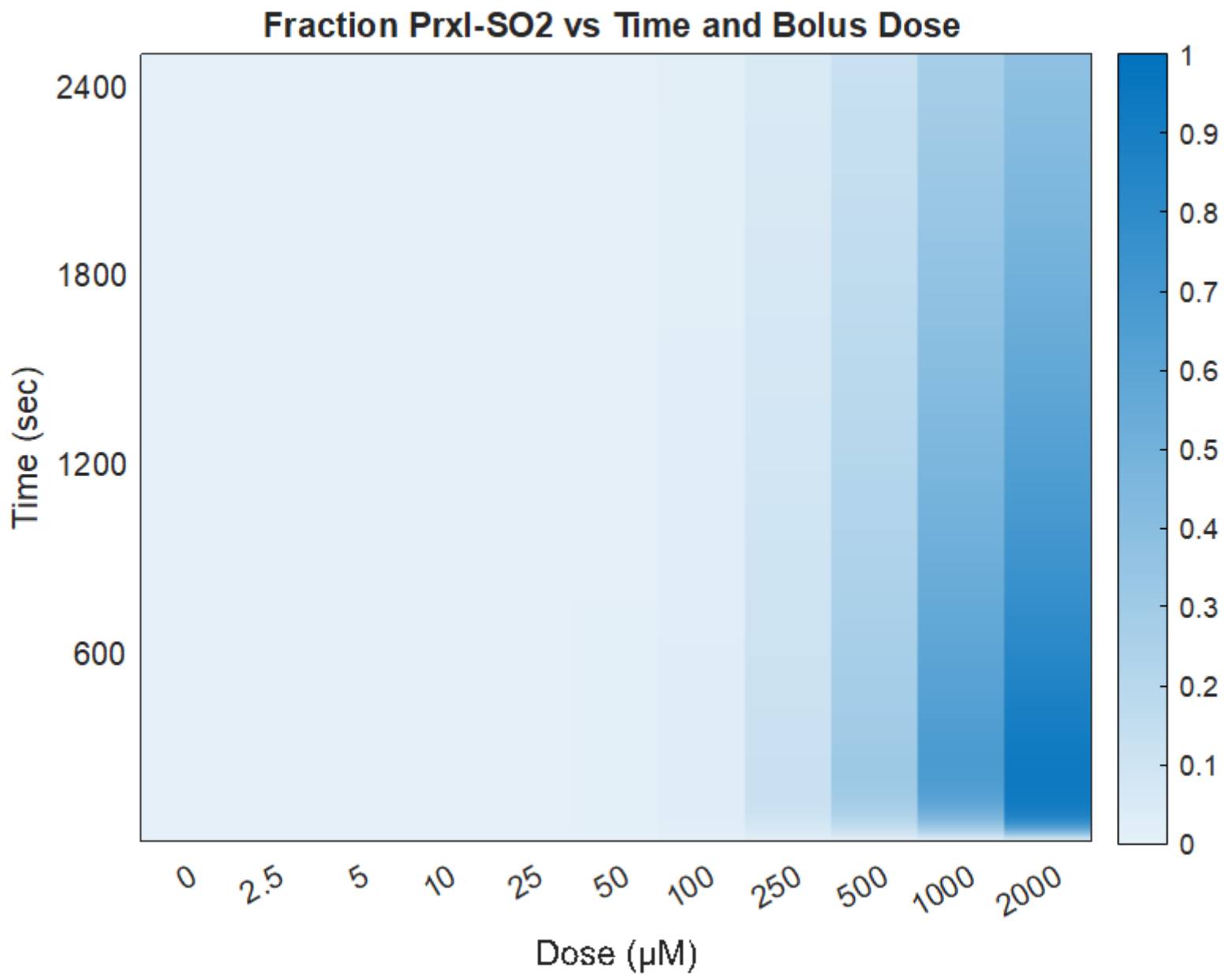

**Figure 5:** Hyperoxidation heatmaps. The fraction of total peroxiredoxin stock that is hyperoxidized is plotted on the color axis, for various timepoints and bolus doses. Technical note regarding the control time-course, labeled 0μM: to prevent mathematical impossibilities interfering with the numerical solution's behavior, the control time course is simulated using a negligible but nonzero dose of 1nM (equivalent to ~0.01nM in the cytoplasm of a single cell).

We were also interested in the nature of the hyperoxidation threshold and, in particular, the curvature of hyperoxidation versus bolus dose. One natural question is whether hyperoxidation grows progressively (e.g., linearly or exponentially) with dosage, or whether it exhibits a sigmoidal, ultrasensitive response. Examining sequential columns for PrxII hyperoxidation in Figure 5 begins to address this question. In the PrxII heatmap (Figure 5a), exactly six (columns) show noteworthy hyperoxidation. For the first three, each increase in dose effects a notable increase in hyperoxidation (darkening of the column). However, increases in dose beyond 250μM cause little to no change in hyperoxidation dynamics; following boluses of 250μM and greater, PrxII hyperoxidation proceeds relatively homogeneously in time. This suggests that peroxiredoxin hyperoxidation dose response is more sigmoidal than linear, exhibiting switchlike dynamics. We examine this further in the next section.

*3.1.2 Sequential, dose-dependent hyperoxidation at a fixed timepoint*

To specifically examine the curvature of peroxiredoxin hyperoxidation versus bolus dose, we next examined dose response data at fixed timepoints post-bolus. Strikingly, the dose-response curves exhibit precisely the hypothesized staggered sigmoidal geometry. For doses below 25μM, both peroxiredoxins remain in the main cycle of the PTRS (i.e., non-hyperoxidized states), but for doses ranging from 25μM to 250μM, PrxII exhibits a sharp sigmoidal increase in hyperoxidation percentage. PrxI exhibits similar curvature, but only at the higher dose range of 250μM to 2000μM, creating three distinct regions in the plot: below hyperoxidation for both peroxiredoxins, between their hyperoxidation thresholds, and above hyperoxidation thresholds for both. As a proof-of-principle, the clear existence of such regions demonstrates that peroxiredoxin hyperoxidation may act as a sensor for oxidative stress severity that can enact three response regimes. Having established these regions, we now return to the aforementioned competing hypotheses that tie hyperoxidation to disulfide signaling (Figure 1).

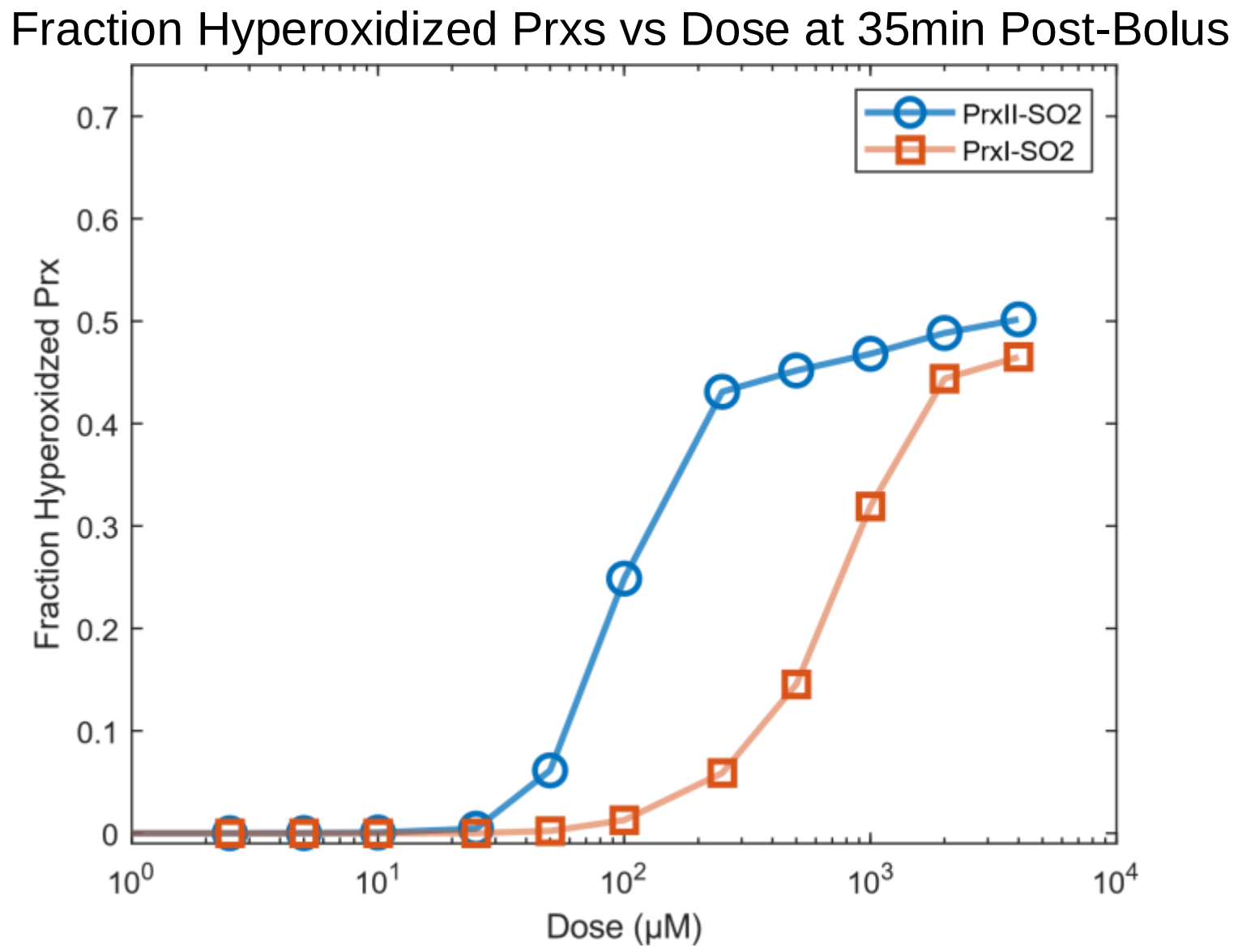

**Figure 6:** Hyperoxidation dose response for both peroxiredoxins. Fractions of the total peroxiredoxin stocks that are hyperoxidized at 35 minutes after bolus addition are plotted versus bolus dose. Note the sigmoidal shapes and the staggering of the curves, which signify separate ultrasensitive switches for hyperoxidation.

Note that the highest hyperoxidation percentages seen depend directly on the choice of fixed timepoint; choosing an earlier timepoint causes the maximum hyperoxidation percentage across dose to rise.

*3.2 Disulfide activity suggests co-activation of PrxI and PrxII at low doses*

A key difference between hypothesis I (isolated peroxiredoxin signaling in dose) and hypothesis II (layered peroxiredoxin signaling in dose) is the overlap of PrxI and PrxII disulfide activity. Hypothesis I proposes that at low doses of $H_2O_2$ exposure (namely those below the PrxII hyperoxidation threshold), PrxII will be the predominant species of peroxiredoxin in the disulfide (signaling) state. Contrastingly, Hypothesis II suggests that in this dose range, both PrxI and PrxII will be in the disulfide state. Following PrxII hyperoxidation, hypothesis I predicts a switch from disulfide activity of PrxII to that of PrxI, whereas hypothesis II predicts continued PrxI-SS activity throughout with PrxII-SS activity disappearing after the PrxII hyperoxidation threshold is reached. To examine these possibilities, we now consider disulfide peroxiredoxin activity for both species versus time across the bolus doses. We intentionally include the temporal dynamics of disulfide prevalence in this analysis so as to avoid the difficulties of comparing signaling (which is relatively transient in nature) at the same timepoint across treatments.

For doses 2.5-100μM, disulfide activity is clearly shared between both peroxiredoxins (Figure 7a-d), providing evidence for hypothesis II in favor of hypothesis I. The scale of this disulfide activity grows with the bolus dose, indicating an intuitive relationship between disulfide activity and signaling strength. Excitingly, the curvature of disulfide activity dose response curves appears to reflect the impact of hyperoxidation (Figure 7d-h). Disulfide activity is short-lived, lasting approximately five minutes, and for doses greater than 25μM, decays logarithmically in the absence of hyperoxidation. However, beginning at 100μM, at which dose PrxII experiences moderate hyperoxidation (Figure 5), the disulfide percentage of PrxII begins to exhibit concavity in the early part of its decay, only transitioning to the convex logarithmic decay after ~100 seconds (Figure 7d). This change in curvature is likely due to hyperoxidation removing available PrxII from the disulfide state before PrxII-SS decays naturally in time. Similarly, beyond doses of 250μM, we see similar inflections in the curvature of PrxI-SS dose responses, ultimately leading to rapid, concave decays in disulfide activity at high $H_2O_2$ doses.

Curiously, the overlapping disulfide activity of both peroxiredoxins does pose a potential contradiction with known biochemistry. Portillo-Ledesma et al. (2018) found that PrxII-SH has significantly higher affinity for $H_2O_2$ than PrxI-SH, as evidenced by a first-oxidation threshold of 4nM as opposed to 0.12μM[21]. This would suggest that there exists a small bolus dose of $H_2O_2$ for which only PrxII-SS becomes active; however, we were unable to find such behavior, instead seeing overlapping disulfide activity of both peroxiredoxins at all low doses, including in simulations not shown (e.g., bolus dose of 1μM). This contradiction may point to a fault of the model or greater complexity in disulfide activation which is not captured. Specifically, Portillo-Ledesma et al. purport similar rate constants for the first oxidation of PrxI and PrxII ($k_{0x}$ in our model) of $1.1 \times 10^8$ and $1.6 \times 10^8$ $M^{-1}s^{-1}$, respectively[21]. However, despite their similar magnitude, these values are achieved at the drastically different oxidation thresholds of 4nM and 0.12μM. Our model parameters do not differentiate $k_{0x}$ for PrxI versus PrxII, which fails to address the difference in requisite $H_2O_2$ dose. Differing second oxidation rate constants in our model parameters may not accurately account for this dose discrepancy, which could explain the observed disulfide dynamics. Nonetheless, the simulated dynamics give rise to the biochemical prediction. Unlike for PrxI/II-SH, affinities for $H_2O_2$ have not been identified for peroxiredoxins in the oxidized sulfenic acid state, PrxI/II-SOH. If PrxII-SOH has a higher affinity for $H_2O_2$ than PrxI-SH, it would follow that hyperoxidation of PrxII should precede disulfide activation of PrxI in dose. Since simulations instead show overlapping disulfide activity at doses below the hyperoxidation threshold for PrxII, we propose the following ranking of $H_2O_2$ affinities instead: PrxII-SH > PrxI-SH > PrxII-SOH > PrxI-SOH. Ideally, this prediction will be tested experimentally in future work.

(a) (b) (c)

(d) (f)

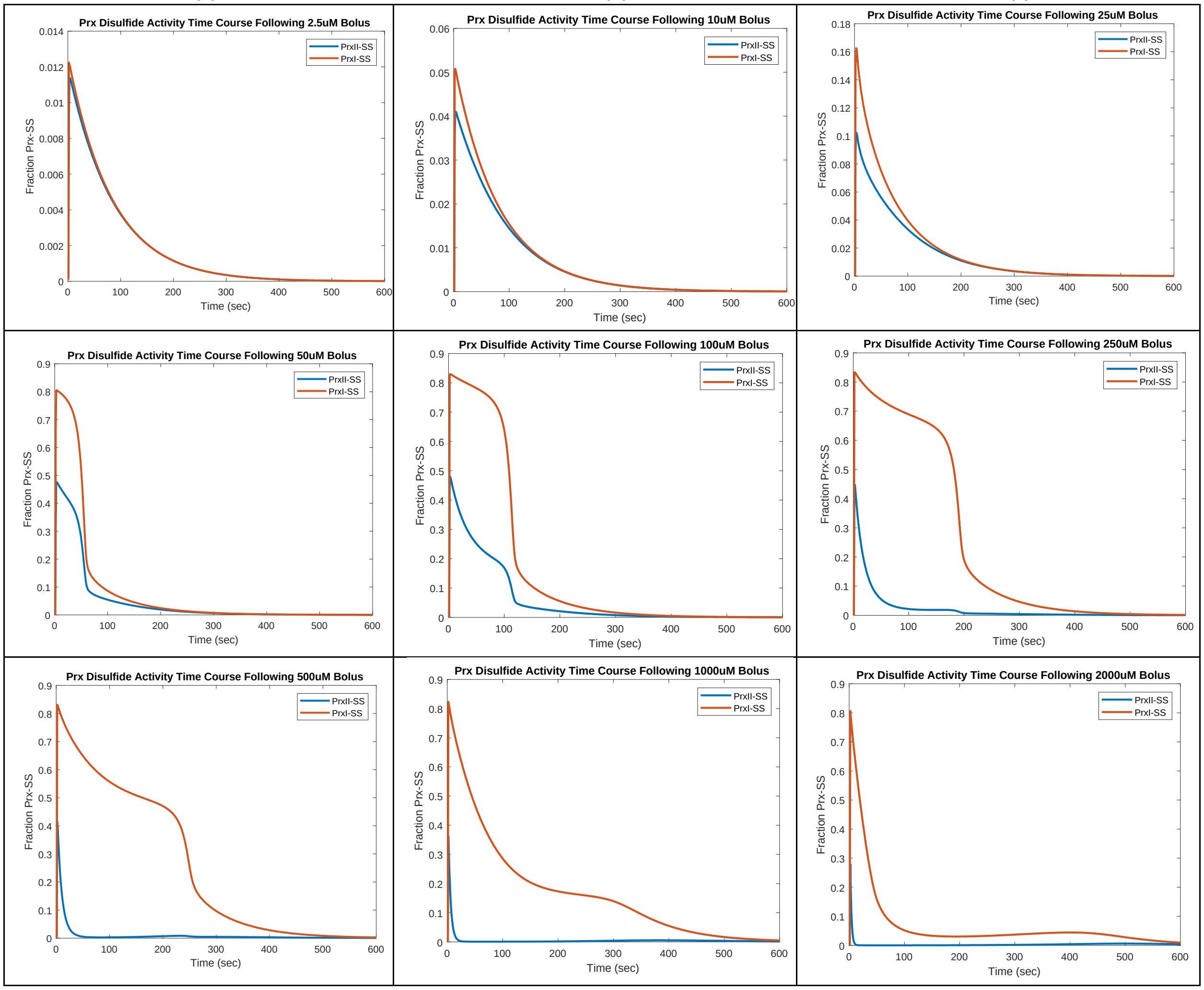

(g) (h) (i)

**Figure 7:** Time courses of disulfide peroxiredoxin activity across doses. Each panel shows the fraction of total peroxiredoxin that is in the disulfide state versus time following a bolus dose. The dose values are: (a) 2.5μM, (b) 10μM, (c) 25μM, (d) 50μM, (e) 100μM, (f) 250 μM, (g) 500μM, (h) 1000μM, (i) 2000μM.

### *3.3 Experimental predictions*

#### *3.3.1 Peroxiredoxin knockouts show asymmetric shifts in hyperoxidation thresholds*

Having established the proof of principle for staggered, sigmoidal hyperoxidation dose response curves, we now turn to making testable predictions based on model simulations. Peroxiredoxin knockouts are of great interest in experimental settings, but the integral role of the PTRS system for healthy cellular function can make these knockouts elusive in practice. We therefore take advantage of the *in silico* setting, simulating peroxiredoxin knockouts by setting the total pool and initial values for all PTRS states for a given peroxiredoxin equal to zero. Excitingly, the hyperoxidation dose response curve for PrxII in the PrxI knockout simulation shifts leftwards, suggesting lower $H_2O_2$ doses will hyperoxidize PrxII (Figure 8a). This agrees with Jose et al.'s (2023) observation that a PrxI knockout leads to activation of the FOXO1 phase at lower $H_2O_2$ doses and less activation of the p53 phase as, in the absence of PrxI, hyperoxidation of PrxII may stand as the only switch between the growth phase and FOXO1 phase.

Surprisingly, performing a similar *in silico* knockout of PrxII does not induce noteworthy changes to the PrxI hyperoxidation dose response. Instead, the resulting PrxI hyperoxidation curve lies almost precisely on top of the WT PrxI curve (Figure 8a). This would suggest that the growth phase of transcriptional activity should be retained even at middling $H_2O_2$ doses, which in WT conditions would have induced hyperoxidation of PrxII. In unpublished work, an *in vitro* knockout of PrxII in MCF7 cells has been seen to be harder to obtain than that for PrxI; as such, future experimental work will provide interesting comparison against this prediction.

To examine potential causes of the lowered PrxII hyperoxidation threshold in the PrxI knockout, we also consider the size of the peroxiredoxin pool in each knockout setting. Since the stock of PrxI far exceeds that of PrxII (PrxITotal = 110 >> PrxIITotal = 32), a PrxI knockout significantly decreases the total available pool of peroxiredoxins to react to oxidative stress. To isolate the effect of this reduction on the observed decrease in PrxII hyperoxidation threshold under a PrxI knockout, we simulated the PrxI knockout setting but with an enlarged stock of PrxII equal to the total pool size (110 + 32 = 142) under wild-type conditions (Figure 8b). Maintaining a WT pool size largely recovers the WT dose response of PrxII hyperoxidation, restoring the inflection point to a dose of ~50μM. Note, however, that some changes in the curvature of the dose response do persist, suggesting that the PrxI knockout does effect changes in PTRS dynamics because of its specific $H_2O_2$ affinity. In a similar fashion, simulating the PrxII knockout setting but with a reduced PrxI stock (equal to the WT PrxII stock, 32) produces a reduction in the PrxI hyperoxidation threshold, where the standard knockout does not (Figure 8c). These findings suggest that the size of the total peroxiredoxin pool may be a driving factor in determining the hyperoxidation thresholds, despite affinity differences between PrxI-SH and PrxII-SH for $H_2O_2$ and rates of condensation (disulfide formation) and hyperoxidation. Moreover, this implication fits with hypothesis II; if transcriptional activity phases are produced by the combinations of disulfide peroxiredoxins that are active at specific doses, then changes in the size of the total pool of peroxiredoxins should influence these phases regardless of preferential $H_2O_2$ affinity of one peroxiredoxin over another.

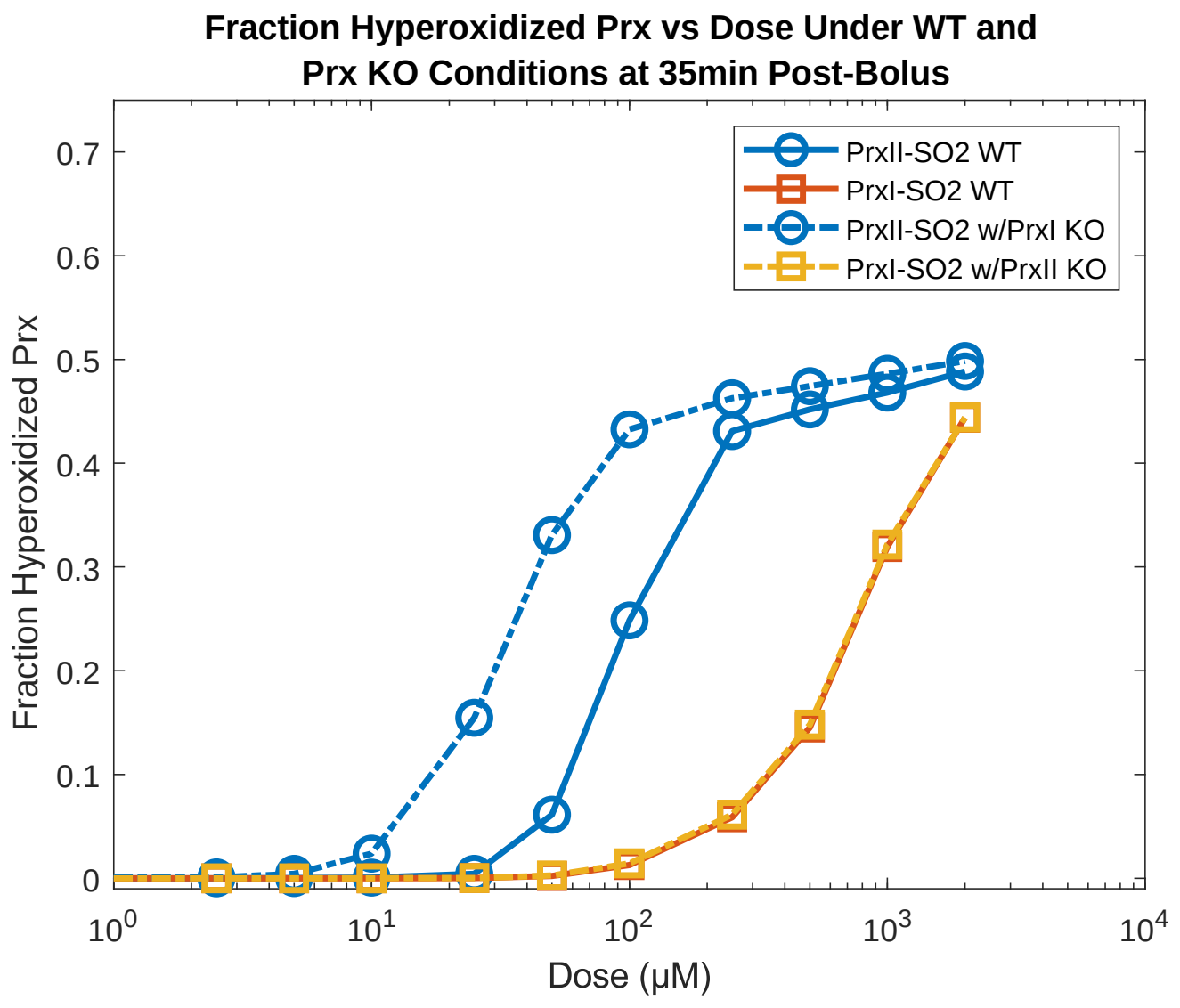

(a)

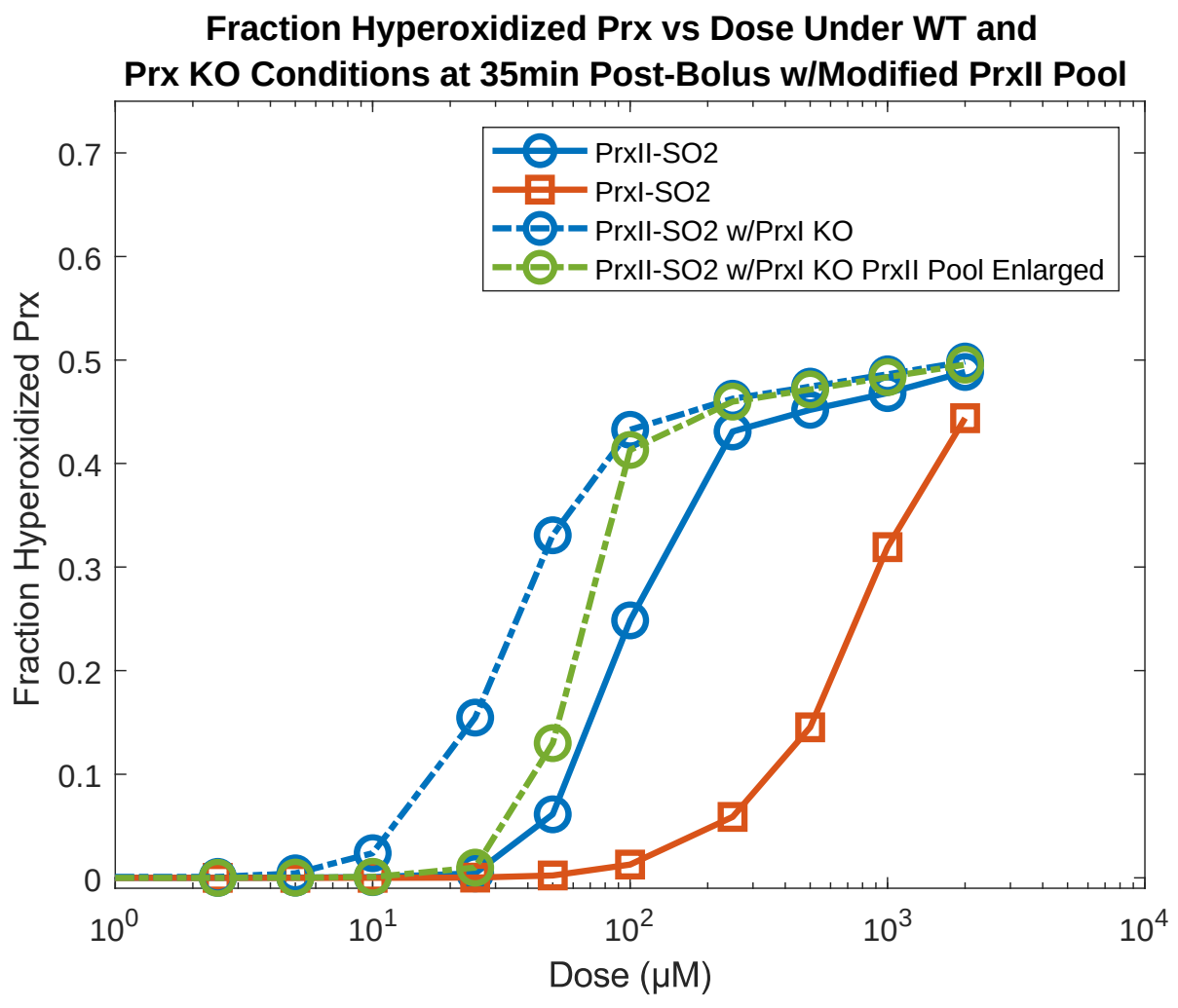

(b)

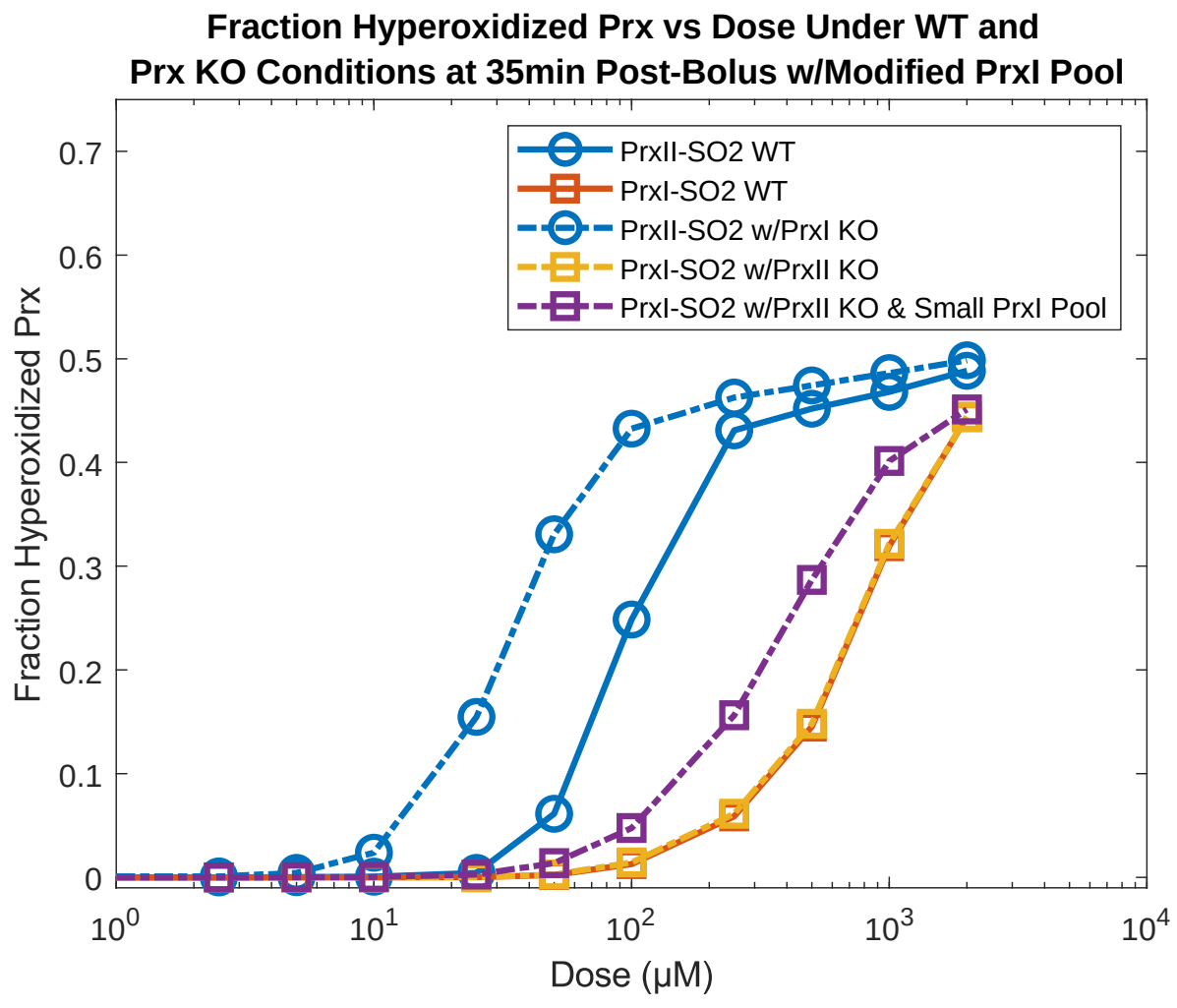

(c)

**Figure 8:** Simulations of the hyperoxidation dose response curves under knockout conditions. Fractions of the total peroxiredoxin stocks that are hyperoxidized at 35 minutes after bolus addition are plotted versus bolus dose. In panel (a), copies of these dose response curves are also plotted under the conditions of a PrxI knockout and those of a PrxII knockout. In panel (b), copies of these dose response curves are also plotted under the conditions of a PrxI knockout and under those of a PrxI knockout with the adjusted parameter PrxIITotal=110. In panel (c), copies of these dose response curves are also plotted under the conditions of a PrxI knockout, under the conditions of a PrxII knockout, and under those of a PrxII knockout with the adjusted parameter PrxITotal=32.

## 4 Discussion

We have provided a proof of principle that the PTRS may direct sequential waves of transcriptional activity in response to oxidative stress by means of hypothesis II. However, this examination of hypotheses was not comprehensive in its comparison nor exhaustive in considering alternatives. As such, it is possible that some combination of hypotheses I and II is more realistic, wherein there are both doses at which peroxiredoxins signal individually as well as in concert. Both hypotheses leave room for debate in the activation link between peroxiredoxins and transcription factors; despite supporting the existence of such links, our work does not shed light on their nature (e.g., direct, indirect). Moreover, in this study we did not examine the role of PrxV, which could be integrated into the PTRS differently than proposed in hypotheses I and II. In addition, our assumption that the disulfide oxidation state is the only signaling state is likely to be an oversimplification, as many other signaling redox mechanisms are known and more still are theoretically feasible. However, the beauty of the mathematical approach is that much can still be done in these regards, both in isolation and as a complement to biological experimentation.

Exemplifying the latter, we produced multiple predictions about the nature of the PTRS (e.g., dynamics under knockout conditions, $H_2O_2$ affinities) which can and should be tested experimentally. Despite our parameter sensitivity analysis and cross-referencing of values, parameters are inevitably hard to precisely determine, and as such leave room for doubt in our model predictions. For instance, although MCF7 cells show similar qualitative dynamics *in silico*, these effects were subdued when compared to those for HEK293 cells, suggesting that the effect of parameters is not one to be neglected.

Future work using this model holds great potential. One testable simulation which is mathematically appealing would be to simulate the effects of TrxR inhibitors, such as Auranofin, by setting the parameter $V_{max}$ to zero. Additionally, further exploration of the effects of the total stocks of each peroxiredoxin could be extended to include the total thioredoxin stock as well. Such experiments would facilitate exploration of the roles of Trx and TrxR, which were relatively untouched in the current work.

**Acknowledgments**

I would like to thank my co-advisors, Andrew Paek and Kevin Lin, both for their insight, and more importantly, their amiability, flexibility, and kindness in meeting with me and accommodating my work style. Additionally, Armindo Salvador enabled this work through tremendously valuable correspondence; without his expertise and generosity, the relevance of this modeling work would be greatly diminished. Moreover, the key insights of this work would not have been found without the valuable conversations I shared with Chance Parkinson, Bryce Wilson, Allison Moreno, Woody March-Steinman, and especially Lisa Jose, whose work laid the foundation for this project.

To all members of the Paek Lab–thank you for your generosity with your time and your friendship. Thanks to Woody for being my math buddy in lab, to Lisa Shanks for sharing stories and jokes that make our side of the lab the better one, and to Lisa Jose, Allison Moreno, and Kathleen Lasick for being mentors more warm, inclusive and encouraging than I could ever have asked for. Finally, I would like to express my continued gratitude for the peerless inspiration that was Julie Huynh; without her, I most likely would not be in this line of work, and if I was, I would most certainly be worse at it.

And of course, it would not be fair to end this work without expressing thanks to my most steadfast source of support: my parents. I love you both.

**References**

[1] Sies, H., and Jones, D.P. (2020). Reactive oxygen species (ROS) as pleiotropic physiological signalling agents. Nat. Rev. Mol. Cell Biol. 2020 217 21, 363–383. https://doi.org/10.1038/s41580-020-0230-3.

[2] Arnold, R.S., Shi, J., Murad, E., Whalen, A.M., Sun, C.Q., Polavarapu, R., Parthasarathy, S., Petros, J.A., and Lambeth, J.D. (2001). Hydrogen peroxide mediates the cell growth and transformation caused by the mitogenic oxidase Nox1. Proc. Natl. Acad. Sci. U. S. A. 98, 5550–5555. https://doi.org/10.1073/pnas.101505898.

[3] Sigaud, S., Evelson, P., and González-Flecha, B. (2004). H2O2-Induced Proliferation of Primary Alveolar Epithelial Cells Is Mediated by MAP Kinases. https://home.liebertpub.com/ars 7, 6–13. https://doi.org/10.1089/ARS.2005.7.6.

[4] Niethammer, P., Grabher, C., Look, A.T., and Mitchison, T.J. (2009). A tissue-scale gradient of hydrogen peroxide mediates rapid wound detection in zebrafish. Nature 459, 996. https://doi.org/10.1038/NATURE08119.

[5] Manford, A.G., Rodríguez-Pérez, F., Shih, K.Y., Shi, Z., Berdan, C.A., Choe, M., Titov, D. V., Nomura, D.K., and Rape, M. (2020). A Cellular Mechanism to Detect and Alleviate Reductive Stress. Cell 183, 46-61.e21. https://doi.org/10.1016/j.cell.2020.08.034

[6] Sena, L.A., and Chandel, N.S. (2012). Physiological roles of mitochondrial reactive oxygen species. Mol. Cell 48, 158–167. https://doi.org/10.1016/j.molcel.2012.09.025.

[7] Stöcker, S., Maurer, M., Ruppert, T., and Dick, T.P. (2018). A role for 2-Cys peroxiredoxins in facilitating cytosolic protein thiol oxidation. Nat. Chem. Biol. 14, 148–155. https://doi.org/10.1038/nchembio.2536.

[8] Netto, L.E., Antunes, F. (2016). The Roles of Peroxiredoxin and Thioredoxin in Hydrogen Peroxide Sensing and in Signal Transduction. Mol Cells. 39 (1), 65-71. https://doi.org/10.14348/molcells.2016.2349.

[9] Thomas, C., Mackey, M.M., Diaz, A.A., and Cox, D.P. (2009). Hydroxyl radical is produced via the Fenton reaction in submitochondrial particles under oxidative stress: Implications for diseases associated with iron accumulation. Redox Rep. 14, 102–108. https://doi.org/10.1179/135100009X392566.

[10] Keyer, K., and Imlay, J.A. (1996). Superoxide accelerates DNA damage by elevating free-iron levels. Proc. Natl. Acad. Sci. U. S. A. 93, 13635–13640. https://doi.org/10.1073/pnas.93.24.13635.

[11] Jose, E., March-Steinman, W., Wilson, B.A., Shanks, L., Paek, A.L. (2023). bioRxiv 2023.03.07.531593. https://doi.org/10.1101/2023.03.07.531593

[12] Liu, D., and Xu, Y. (2011). P53, oxidative stress, and aging. Antioxidants Redox Signal. 15, 1669–1678. https://doi.org/10.1089/ars.2010.3644.

[13] Madan, E., Gogna, R., Bhatt, M., Pati, U., Kuppusamy, P., and Mahdi, A.A. (2011). Regulation of glucose metabolism by p53: Emerging new roles for the tumor suppressor. Oncotarget 2, 948–957. https://doi.org/10.18632/oncotarget.389.

[14] Klotz, L.O., Sánchez-Ramos, C., Prieto-Arroyo, I., Urbánek, P., Steinbrenner, H., and Monsalve, M. (2015). Redox regulation of FoxO transcription factors. Redox Biol. 6, 51. https://doi.org/10.1016/J.REDOX.2015.06.019.

[15] Shi, T., and Dansen, T.B. (2020). Reactive Oxygen Species Induced p53 Activation: DNA Damage, Redox Signaling, or Both? Antioxidants Redox Signal. 33, 839–859. https://doi.org/10.1089/ars.2020.8074.

[16] Ma, Q. (2013). Role of Nrf2 in oxidative stress and toxicity. Annu. Rev. Pharmacol. Toxicol. 53, 401–426. https://doi.org/10.1146/annurev-pharmtox-011112-140320.

[17] Ahn, S.G., and Thiele, D.J. (2003). Redox regulation of mammalian heat shock factor 1 is essential for Hsp gene activation and protection from stress. Genes Dev. 17, 516–528. https://doi.org/10.1101/gad.1044503.

[18] Lin, T.-Y., Cantley, L.C., and DeNicola, G.M. (2016). NRF2 Rewires Cellular Metabolism to Support the Antioxidant Response. In A Master Regulator of Oxidative Stress - The Transcription Factor Nrf2 (InTech). https://doi.org/10.5772/65141.

[19] Shi, T., van Soest, D.M.K., Polderman, P.E., Burgering, B.M.T., and Dansen, T.B. (2021). DNA damage and oxidant stress activate p53 through differential upstream signaling pathways. Free Radic. Biol. Med. 172, 298–311. https://doi.org/10.1016/J.FREERADBIOMED.2021.06.013.

[20] Lehtinen, M.K., Yuan, Z., Boag, P.R., Yang, Y., Villén, J., Becker, E.B.E., DiBacco, S., de la Iglesia, N., Gygi, S., Blackwell, T.K., et al. (2006). A Conserved MST-FOXO Signaling Pathway Mediates Oxidative-Stress Responses and Extends Life Span. Cell 125, 987–1001. https://doi.org/10.1016/j.cell.2006.03.046.

[21] Portillo-Ledesma, S., Randall, L.M., Parsonage, D., Dalla Rizza, J., Karplus, P.A., Poole, L.B., Denicola, A., Ferrer-Sueta, G. (2018). Differential Kinetics of Two-Cysteine Peroxiredoxin Disulfide Formation Reveal a Novel Model for Peroxide Sensing. Biochemistry. 57 (24), 3416-3424. https://doi.org/10.1021/acs.biochem.8b00188.

[22] van Dam, L., Pagès-Gallego, M., Polderman, P.E., van Es, R.M., Burgering, B.M.T., Vos, H.R., Dansen, T.B. (2021). The Human 2-Cys Peroxiredoxins form Widespread, Cysteine-Dependent- and Isoform-Specific Protein-Protein Interactions. Antioxidants (Basel). 10 (4), 627. https://doi.org/10.3390/antiox10040627.

[23] Bolduc, J.A., Nelson, K.J., Haynes, A.C., Lee, J., Reisz, J.A., Graff, A.H., Clodfelter, J.E., Parsonage, D., Poole, L.B., Furdui, C.M., Lowther, W.T. (2018). Novel hyperoxidation resistance motifs in 2-Cys peroxiredoxins. J Biol Chem. 293 (30), 11901-11912. https://doi.org/10.1074/jbc.RA117.001690.

[24] Selvaggio, G., Coelho, P.M.B.M., Salvador, A. (2018). Mapping the phenotypic repertoire of the cytoplasmic 2-Cys peroxiredoxin - Thioredoxin system. 1. Understanding commonalities and differences among cell types. Redox Biol. 15, 297-315. https://doi.org/10.1016/j.redox.2017.12.008.

[25] Antunes, F., Brito, P.M. (2017). Quantitative biology of hydrogen peroxide signaling. Redox Biol. 13, 1-7. https://doi.org/10.1016/j.redox.2017.04.039.

[26] Benfeitas, R., Selvaggio, G., Antunes, F., Coelho, P.M., Salvador, A. (2014). Hydrogen peroxide metabolism and sensing in human erythrocytes: a validated kinetic model and reappraisal of the role of peroxiredoxin II. Free Radic Biol Med. 74, 35-49. https://doi.org/10.1016/j.freeradbiomed.2014.06.007.

[27] Sobotta, M.C., Barata, A.G., Schmidt, U., Mueller, S., Millonig, G., Dick, TP. (2013). Exposing cells to H2O2: a quantitative comparison between continuous low-dose and one-time high-dose treatments. Free Radic Biol Med. 60, 325-35. https://doi.org/10.1016/j.freeradbiomed.2013.02.017.

[28] "Diameter of Hek-293 Cell." Edited by Uri M, Diameter of HEK-293 Cell - BNID 108893, Harvard University, https://bionumbers.hms.harvard.edu/bionumber.aspx?id=108893&ver=3&trm=HEK%2B293&org=.

[29] Wagner, B.A., Venkataraman, S., Buettner, G.R. (2011). The rate of oxygen utilization by cells. Free Radic Biol Med. 51 (3), 700-12. https://doi.org/10.1016/j.freeradbiomed.2011.05.024 (p.707 right column bottom paragraph; obtained from Bionumbers, https://bionumbers.hms.harvard.edu/bionumber.aspx?id=115154&ver=1).

[30] Moon, H.S., Kwon, K., Hyun, K.A., Seok Sim, T., Chan Park, J., Lee, J.G., Jung, H.I. (2013). Continual collection and re-separation of circulating tumor cells from blood using multi-stage multi-orifice flow fractionation. Biomicrofluidics. 7 (1), 14105. https://doi.org/10.1063/1.4788914.